\documentclass{emulateapj}

\usepackage{epsfig,amsmath,amssymb,amsfonts,mathrsfs,latexsym,graphicx,lscape}
\usepackage{hyperref}

\begin{document}

\title{Low Frequency (11 mHz) Oscillations in H1743--322: A New Class
of Black Hole QPOs ?}

\author{D. Altamirano\altaffilmark{1} \& T. Strohmayer\altaffilmark{2}}

\altaffiltext{1}{Email: d.altamirano@uva.nl ; Astronomical Institute,
  ``Anton Pannekoek'', University of Amsterdam, Science Park 904,
  1098XH, Amsterdam, The Netherlands}

\altaffiltext{2}{Astrophysics Science Division, Mail Code 662, NASA
Goddard Space Flight Center, Greenbelt, MD 20771, U.S.A.}

\begin{abstract}

We report the discovery of quasi-periodic oscillations (QPO) at
$\sim$11 mHz in two RXTE observations and one Chandra observation of
the black hole candidate H1743--322.
The QPO is observed only at the beginning of the 2010 and 2011
outbursts at similar hard color and intensity, suggestive of an
accretion state dependence for the QPO.
Although its frequency appears to be correlated with X-ray intensity on
timescales of a day, in successive outbursts eight months apart we
measure a QPO frequency that differs by less than $\approx$2.2 mHz 
while the intensity had changed significantly.
We show that this $\sim$11 mHz QPO is different from the so-called
Type-C QPOs seen in black holes and that the mechanisms that produce
the two flavors of variability are most probably independent.
After comparing this QPO with other variability phenomena seen in
accreting black holes and neutron stars, we conclude that it best
resembles the so-called ``1 Hz'' QPOs seen in dipping neutron star
systems, although having a significantly lower (1-2 orders of
magnitude) frequency. If confirmed, H1743--322 is the first black hole
showing this type of variability.
Given the unusual characteristics and the hard-state dependence of the
$\sim$11~mHz QPO, we also speculate whether these oscillations could
instead be related to the radio jets observed in H1743--322.
A systematic search for this type of low-frequency QPOs in similar
systems is needed to test this speculation.
In any case, it remains unexplained why these QPOs have only been seen
in the last two outbursts of H1743--322.

\end{abstract}
\keywords{ X-rays: binaries --- binaries: close
  --- stars: individual (H1743--322) --- Black hole
  physics}

\section{Introduction}\label{sec:intro}

Quasi-periodic oscillations (QPOs) with characteristic frequencies
between $\sim$1 mHz and hundreds of Hz have now been observed in the
X-ray flux of many low-mass X-ray binaries (LMXBs) containing neutron
stars (NSs) and black holes (BHs).
The characteristics of many of these QPOs are known to generally
correlate with the source spectral states and/or X-ray luminosity
\citep[see, e.g.,][for a review]{Vanderklis06}.

A number of QPOs and broad band variability components can be present
simultaneously in the power spectra of the X-ray light curves of 
accreting NS systems. These can be modeled with Lorentzian
components and are often denoted by their characteristic frequencies
\citep[examples include the so-called break, the hump, the LF QPO, the
\textit{$\ell$ow} component and the upper and lower kilohertz QPOs,
e.g.,][]{Altamirano08}.
Their frequencies are known to be correlated to each other
\citep{Wijnands99a,Belloni02}.
However, there are also components which appear to be uncorrelated.
Among these uncorrelated exceptions are: (i) the hecto-Hz QPOs which
have frequencies constrained within $\sim$100~Hz and $\sim$300~Hz
\citep[e.g.,][]{Altamirano08}, (ii) the 1~Hz QPOs in dipping sources
\citep[e.g.,][]{Jonker99, Jonker00, Homan03a}, (iii) a QPO with
frequencies in the mHz range which very likely results from marginally
stable nuclear burning of hydrogen/helium on the NS surface
\citep[e.g.,][]{Heger07} and (iv) the 1~Hz flaring observed in two
accreting millisecond X-ray pulsars when the sources were observed at
low-luminosities \citep[e.g.][]{Wijnands04,Patruno10b}.

BHs generally show three main types of low-frequency QPOs \citep[Types
A, B and C, e.g.,][]{Casella05} and, in a few cases,
high-frequency QPOs \citep[with frequencies between 70~Hz and 450~Hz,
e.g.,][for a review]{Vanderklis06}.  
In some cases observed QPOs are not easily identified with either Type
A, B or C; this may be due to having data of low statistical quality
or a source being observed in an unusual spectral state. In general,
some of the same variability components appear to be present in both
NS and BHs \citep{Miyamoto93,Vanderklis94,Belloni02}. For example,
\citealt{Casella05} argue that the Type A, B and C QPOs in black holes
correspond to the so-called Flaring Branch, Normal Branch and
Horizontal Branch Oscillations, respectively, that were identified in
high luminosity NS systems (the ``Z'' sources).

Outside of the Type A, B, and C QPO classifications there is additional
low-frequency variability in BHs, including for example the
``heart-beat'' QPOs, observed so far only in the BHs GRS~1915+105
\citep[e.g, ][]{Belloni00} and IGR~J17091--3624
\citep[e.g.,][]{Altamirano11d}.

Recently, \citet{Strohmayer11a} reported on a \textit{Rossi X-ray
Timing Explorer} (RXTE) observation of the 2011 outburst of the black
hole candidate (BHC) H1743--322. In addition to the typical Type-C QPO
and strong broad-band noise generally seen at the beginning of the
outbursts of BH systems, \citet{Strohmayer11a} found an unusual
$\sim$91s QPO which they suggested might be due to a second active
source in the $1^o$ Proportional Counter Array \citep[PCA;
][]{Jahoda06} field of view (FoV). However, subsequent observations
with {\it Swift} and {\it INTEGRAL} did not show any additional nearby
sources.
Triggered by the possibility that this $\sim$91s QPO might instead be
intrinsic to H1743--322, we searched RXTE, Swift and {\it Chandra}
data for other instances of similar QPOs. In this Letter we report the
discovery of a similar QPO in additional {\it RXTE} and {\it Chandra}
observations of H1743--322 and discuss the implications of our
findings.

H1743--322 was first detected in outburst in 1977
\citep{Kaluzienski77} and then rediscovered by {\it INTEGRAL} in 2003
\citep{Revnivtsev03b}. Since 2003 several smaller outbursts of
H1743--322 have been observed \cite[see][and references
therein]{Motta10}.
From its similarities with the dynamically confirmed BH X-ray binary
XTE J1550-564, the source was classified as a BH candidate \citep[but
  also see \citealt{Homan05b}]{McClintock09}. H1743--22 is at a
distance of $8.5\pm0.8$~kpc \citep{Steiner11} and, from the evidence
of X-ray dipping behavior in one observation \citep{Homan05b}, the
inclination angle of its accretion disc is believed to be relatively
high ($>70^o$) to our line of sight.
However, this high inclination needs to be confirmed, as no thorough
analysis of the dipping behavior has yet been reported.

\begin{figure*} 
\centering
\resizebox{2\columnwidth}{!}{\rotatebox{0}{\includegraphics[clip]{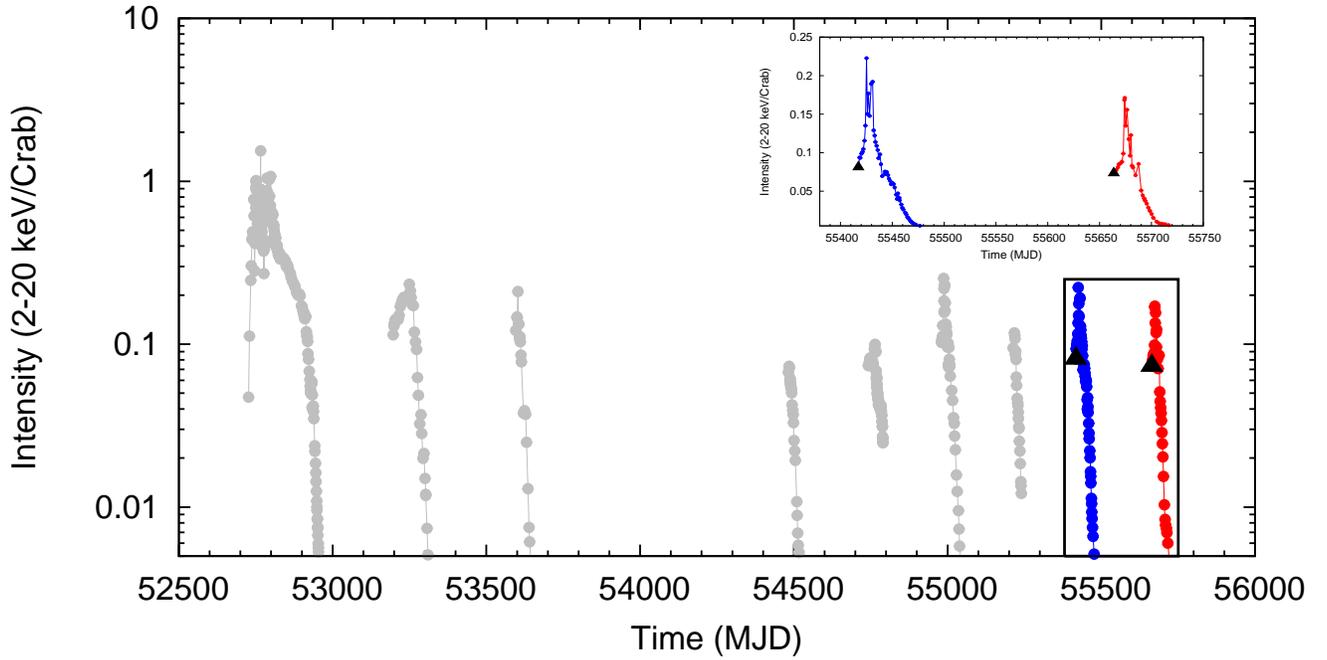}}}
\caption{Crab-normalized light curves of all outbursts of the BHC
  H1743--322 as seen with RXTE pointed observations. Each point
  corresponds to an average per observation. The inset shows the last
  two outbursts. The triangles mark the times when we detected the mHz
  QPOs.}
\label{fig:lc}
\end{figure*}

\begin{figure} 
\centering
\resizebox{1\columnwidth}{!}{\rotatebox{0}{\includegraphics[clip]{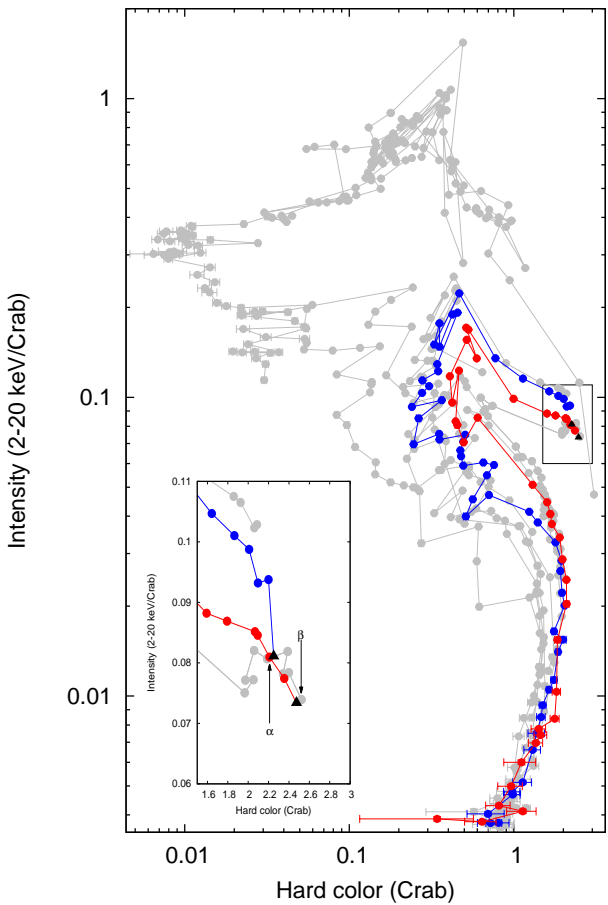}}}
\caption{Hardness--intensity diagram of all outbursts of the BHC
  H1743--322 as seen with RXTE pointed observations. Each point
  corresponds to an average per observation. Inset: blow-up of the
  region in which we detected the mHz QPOs, with triangles denoting
  the mHz QPO observations. Arrows $\alpha$ and $\beta$ mark the
  comparison power spectra in Figure~\ref{fig:pds}. }
\label{fig:hid}
\end{figure}

\begin{figure} 
\centering
\resizebox{1\columnwidth}{!}{\rotatebox{0}{\includegraphics[clip]{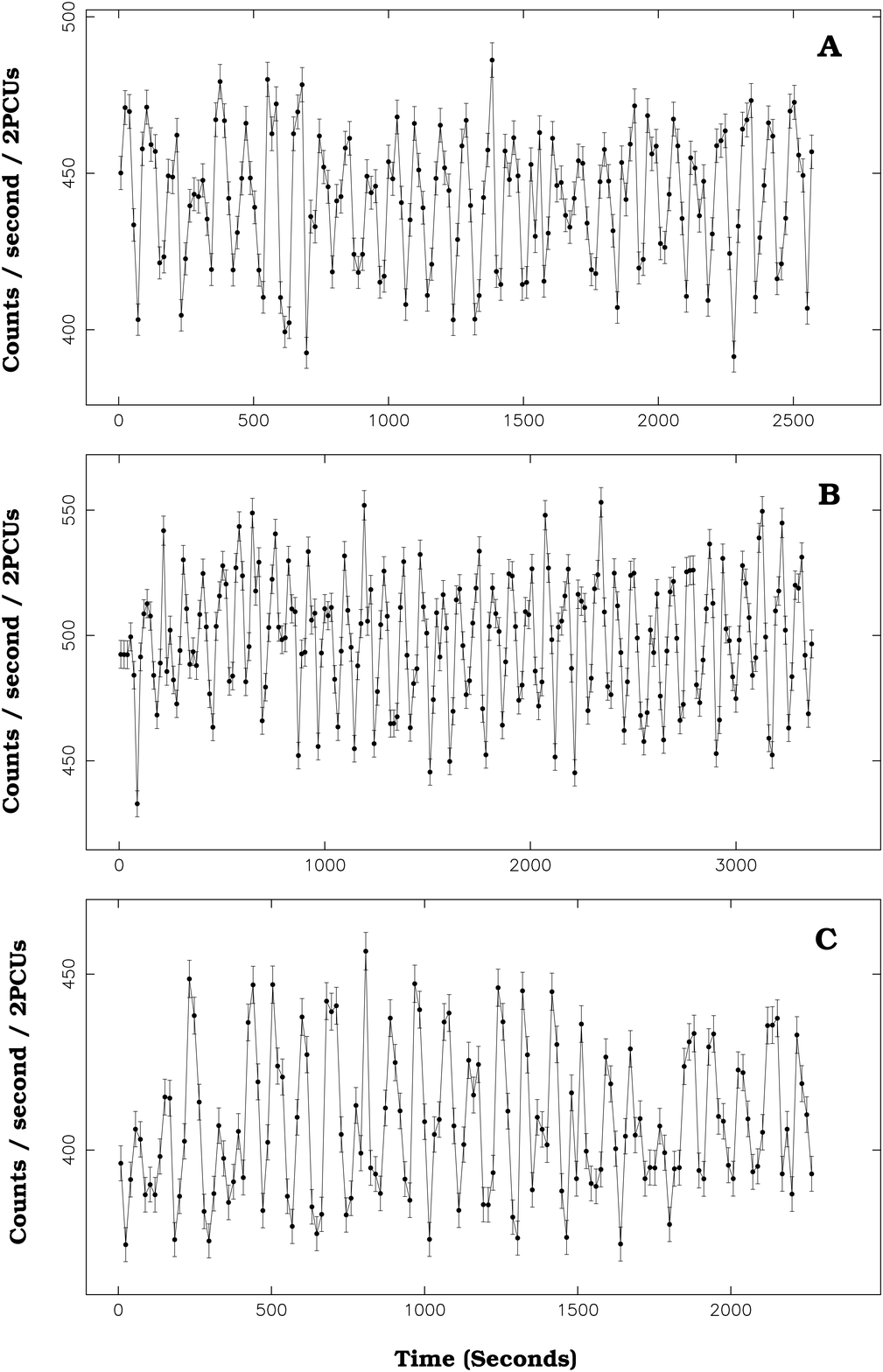}}}
\caption{10s binned RXTE light curves (2--60 keV) of the 3 orbits
  in which we detect the mHz QPOs. Panels A and B correspond to two
  orbits in observation 95368-01-01-00. Panel C corresponds to the
  single orbit observation 96425-01-01-00. }
\label{fig:mhzlc}
\end{figure}

\begin{figure*} 
\centering
\resizebox{2.\columnwidth}{!}{\rotatebox{0}{\includegraphics[clip]{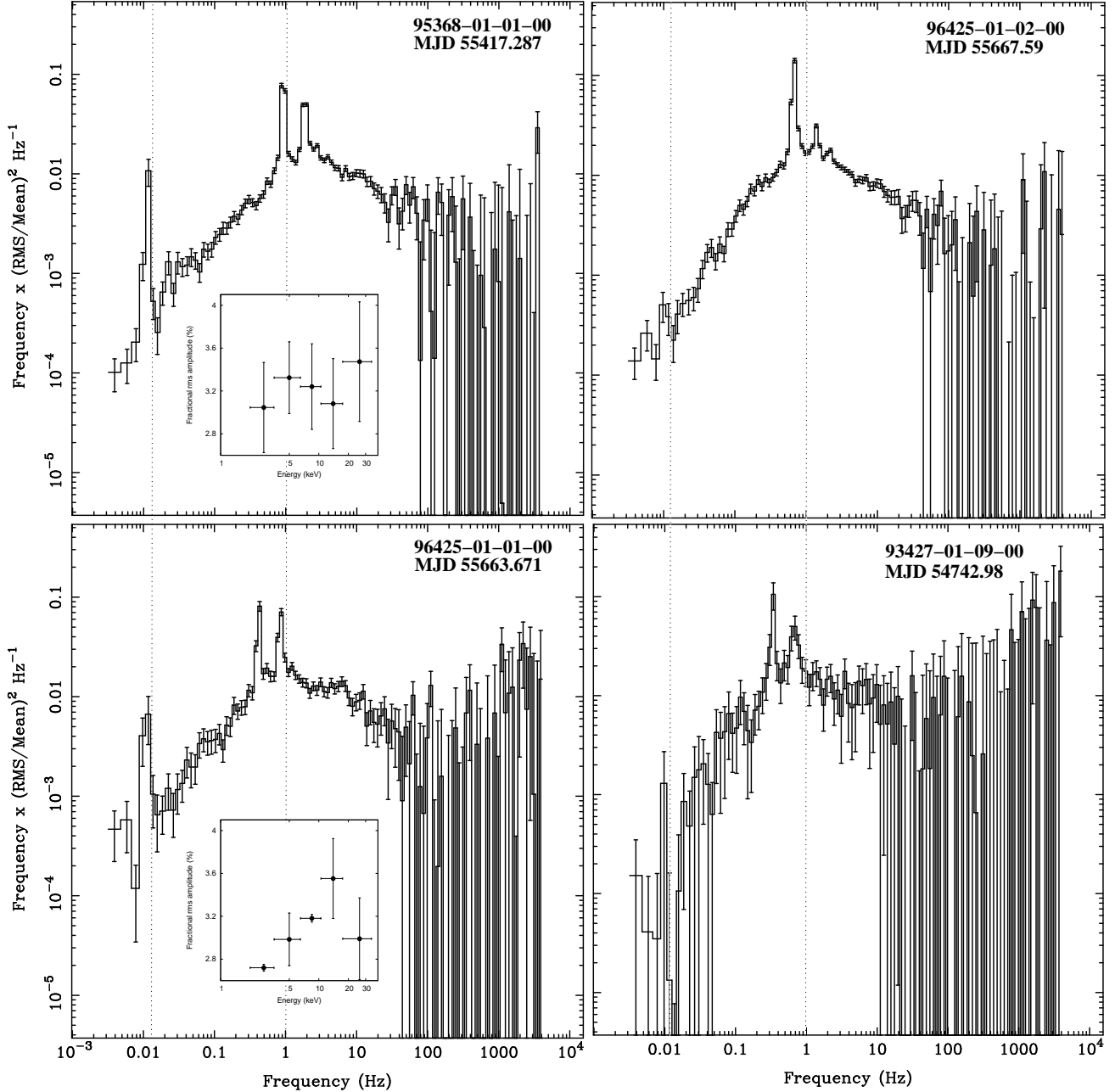}}}
\caption{\textit{Left}: Average power spectra of the two RXTE
  observations where we detect the mHz QPOs.  In the upper panel (2010
  observation) the Type-C QPOs are at $0.919\pm0.001$~Hz and
  $1.852\pm0.001$~Hz while in the lower panel (2011 observation) they
  are at $0.424\pm0.003$~Hz and $0.856\pm0.004$~Hz.  Their quality
  factors (Q) are $13.0\pm0.5$, $10.0\pm0.6$, $11\pm2$ and $9\pm1$ and
  their fractional rms amplitudes $11.9\pm0.2$\%, $9.4\pm0.2$\%
  $10.7\pm0.4$\% $10.9\pm0.5$\%, respectively. The inset panels show
  the fractional rms amplitude vs. energy for the 11 mHz
  QPOs. \textit{Right}: Comparison power spectra without mHz QPOs but
  at similar spectral state (see Figure~\ref{fig:hid} and
  Section~\ref{sec:results}). The vertical dotted lines drawn at 12.5
  mHz and 1 Hz are shown to guide the eye.}
\label{fig:pds}
\end{figure*}

\begin{figure} 
\centering
\resizebox{1\columnwidth}{!}{\rotatebox{0}{\includegraphics[clip]{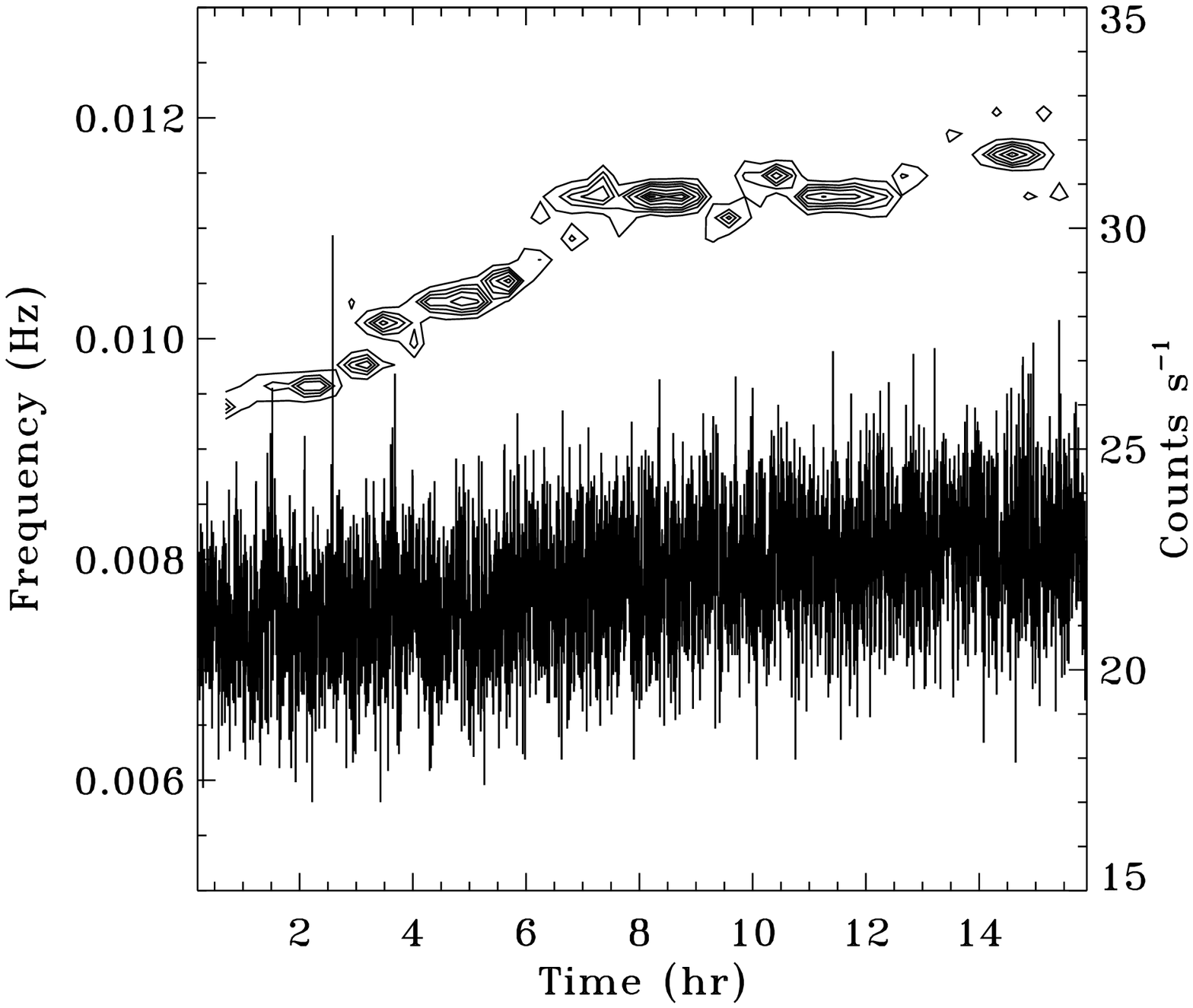}}}
\caption{Dynamical power spectra from the dispersed (i.e., excluding
  the zeroth order image) HETG {\it Chandra} data showing the
  frequency evolution of the mHz QPO. We computed overlapping power
  spectra by using 5000 s intervals, with a new interval beginning
  every 1000 s. Contours of constant Leahy-normalized power are drawn
  at values of 25, 40, 50, 60, 75, 90, 100 and 110. Start time is
  August 8th, 2010, 23:03:48UT.}
\label{fig:dyn}
\end{figure}


\section{Observations and data analysis}\label{sec:obs}

The BHC H1743--322 was observed with the PCA on-board RXTE for a
total of 558 pointed observations, sampling 9 different outbursts
which occurred between March 2003 and June 2011.
Some of the H1743--322 outbursts were also sampled with the Swift/XRT
instrument \citep{Barthelmy05}; we used all the 52 archival XRT
observations to look for mHz QPOs.
We also used a 60ks Chandra observation (id number, 401083), which
was contemporaneous with RXTE observation 95368-01-01-00 (see
Section~\ref{sec:results}). We extracted events from the 0th and 1st order
High Energy Transmission Grating (HETG/ACIS-S) mode data with CIAO 4.3
following standard
recipes\footnote{http://cxc.harvard.edu/ciao/threads/}.

We used the 16s time-resolution PCA Standard 2 mode data to calculate
X-ray colors following the procedure described in
\citet{Altamirano08}. We define hard color as the (16.0--20.0
keV)/(2.0--6.0 keV) count rate ratio and, intensity, as the count rate
in the 2.0--20 keV band. All values have been normalized to the Crab
Nebula on a per PCU basis and are presented here as averages per
observation.
In Figures~\ref{fig:lc} and \ref{fig:hid} we show the light curves and
hardness-intensity diagrams (HIDs) of the nine outbursts observed with
RXTE.

We used 1-sec resolution PCA (2--60~keV; observed in Event, Good-Xenon
and Single Bit modes) and Swift/XRT (0.5--10 keV) light curves to
search for variability on timescales of a fraction of a minute or
longer; this was done on a per orbit basis using Lomb-Scargle
periodograms \citep{Lomb76,Scargle82,Numerical-Recipes}.

To study subsecond variability, we constructed Leahy-normalized power
spectra per RXTE observation.
No background or deadtime corrections were made prior to their
calculation.
We subtracted a predicted dead-time modified Poisson noise spectrum
estimated from the power at frequencies higher than 1500~Hz, where
neither intrinsic noise nor QPOs are known to be present, using the
method developed by \citet{Kleinwolt04}.
The resulting power spectra were converted to squared fractional rms
amplitude \citep{Vanderklis95b}.


\section{Results}\label{sec:results}

We visually inspected each 1s light curve and each of the Lomb-Scargle
periodograms from PCA data and consistently find clear evidence of mHz
QPOs at $\sim$11~mHz in the two observations 95368-01-01-00 (MJD
55417.24, two orbits starting on August 9th, 2010 at 16:05 UT,
averaged periodogram peaks at $\sim$11.4 mHz)
and 96425-01-01-00 (MJD 55663.67, single orbit on April 12th, 2011 at
6:45 UT, averaged periodogram peaks at $\sim$11.1 mHz).
The light curves are shown in Figures~\ref{fig:mhzlc}.
For these two observations we calculated power spectra using 512s data
segments. 
The left panels of Figure~\ref{fig:pds} show the clear $\sim$11 mHz peak
in both cases.
Their averaged quality factor Q is as high as $\sim$100 in the 2010
observation, but also as low as $\sim$10 in the 2011 observation.
The average fractional rms amplitude in the 2--60 keV is
$3.1\pm0.4$~\%. Insets in Figure~\ref{fig:pds} show the 11 mHz QPO
amplitude-energy dependence. 
In the 2010 observation the rms amplitude does not vary with energy,
while in the 2011 observation it first increases and then shows
evidence of a decrease with energy. However, the increase is only
moderate ($<1$\% rms) as compared with other QPO variability, which
generally shows a more substantial increase with energy
\citep[e.g.,][]{Casella04}.

The power spectra in Figure~\ref{fig:pds} also show the typical Type-C
QPOs on top of strong ($\sim$30\% rms amplitude) broad band noise.
Although the frequency of the mHz QPOs is similar in both
observations, the frequency of the Type-C QPOs differs by a factor of
$\sim$2.
We compared the overall power spectra produced using data from
different phases of the 11 mHz QPO, but found no significant changes,
implying that the mechanisms that produce the mHz  and Type-C QPOs
are not closely related.

Figure~\ref{fig:lc} and ~\ref{fig:hid} show that the mHz QPOs were
detected during observations at the beginning of the last two
outbursts, i.e. of the 2010 and 2011 outbursts.
These last two outbursts show very similar tracks in the HID, and the
mHz QPOs appear at a very similar hard color. However, other
observations at similar intensity and hard color do not show evidence
of mHz QPOs.
For comparison, the right panels of Figure~\ref{fig:pds} show the
power spectra of the observations closest in the HID to those where we
detected the mHz QPOs (marked with arrows in the inset of
Figure~\ref{fig:hid}).

Visual inspection of the Swift/XRT data revealed no evidence of mHz
QPOs.
No Swift--RXTE simultaneous observations are available when the mHz
QPOs were detected.
Swift observations occurred approximately two days before and five
days after the 2010 RXTE detection of mHz QPOs, and only 5 days after
the 2011 RXTE detection.

A Chandra/HETG observation started on August 8th, 2010, at 23hs UT and
lasted until approximately 15hs 40min UT of the next day (i.e., until
$\sim20$ min before the RXTE observation 95368-01-01-00 began).
We extracted events both from the zeroth (undispersed) and dispersed
orders and found in both cases a clear QPO at $\sim$11 mHz. In
Figure~\ref{fig:dyn} we show the dynamical power spectrum computed
from the dispersed photons. The QPO frequency drifts upward with time
and appears to be positively correlated with the intensity.  The
presence of the QPO signal in the zeroth order photons at a sky
position consistent with the known position of H1743--322 confirms
that the mHz QPO is intrinsic to this source, and not from a hitherto
unidentified nearby source. A phase-resolved spectral analysis of
these data is beyond the scope of this paper and will be reported
elsewhere.

%


\section{Discussion}

We report the discovery of QPOs at $\sim$11 mHz
in two RXTE and one Chandra observation of the BHC H1743--322. In
successive outbursts eight months apart we measure a QPO frequency
that differs by less than $\approx1.5$ mHz (including the 60ks
duration of the {\it Chandra} observation, see Figure~\ref{fig:dyn}).
The fractional rms amplitude of the oscillation appears to be
correlated with energy in the 2011 observation, but consistent with a
constant value in the 2010 observation.
The QPO is observed at the beginning of two different outbursts at
similar hard color and intensity, suggestive of an accretion state
dependence for the QPO.
Although the {\it Chandra} data reveal that the QPO frequency might be
correlated with intensity on timescales of hours, this correlation
probably changes in between outbursts, as we find the same frequency
(within $\approx0.4$ mHz) in observations separated by about 800 days
and at source intensities different by $\approx10$ mCrab \citep[this
  resembles the so called ``parallel tracks'' observed in the
  frequency vs. intensity diagrams of NS kHz QPOs, e.g,
][]{Mendez99a}.
Except for the 11 mHz QPOs, the RXTE power spectra of these two
observations are typical of the low-hard state of BH LMXBs, showing
Type-C QPOs on top of strong broad band noise.
Given that 
(i) the power spectra characteristics do not change with mHz QPO phase
and
(ii) the frequency of the mHz QPOs is rather constant
($\lesssim0.4$mHz) between the 2010 and 2011 RXTE observation, while
the Type-C QPO frequency varies by a factor of about 2,
we conclude that mechanisms that produce the mHz QPOs and Type C QPOs
are not closely related.

The fact that the frequency of these new oscillations is fairly
constant raises the question of whether they represent a
characteristic frequency of a process not yet identified before.
Several types of QPOs with frequencies in the mHz range have been
reported in two BH systems and in some NS systems.
Below we compare our results with those seen in other sources and
discuss whether we can identify the 11~mHz QPOs with any of them based
on the characteristics of the oscillations and the source state in
which they occur.

%
%

Highly structured, high-amplitude variability has been
seen in the BH systems GRS~1915+105 \citep[e.g.,][]{Belloni00} and
IGR~J17091--3624 \citep[e.g.,][]{Altamirano11d}. Some of these
variations are known as ``heartbeats'' and are thought to be due to
limit cycles of accretion and ejection in an unstable disk
\citep[e.g.,][]{Neilsen11}.
These ``heartbeat'' QPOs are in the mHz range, occur only during the
high-luminosity, soft-state of these two BH systems and, at least in
the case of IGR~J17091--3624, can have rms amplitudes as low as
$\sim$3\% \citep[e.g.,][]{Altamirano11d}.
For H1743--322 we only find the new mHz QPOs during the rise of the
outburst \citep[at $L_X < 3 \cdot 10^{37}$ egs s$^{-1}$,
  e.g.,][]{Motta10}, when the spectrum of the source is dominated by
the hard component. GRS~1915+105 is thought to be very often at an
Eddington or a super-Eddington luminosity \citep[e.g., ][]{Done04};
this could also be the case of IGR~J17091--3624 although the distance
to this source is not yet known \citep[see discussion in
][]{Altamirano11d}.
Given the major differences between source state and luminosity in
GRS~1915+105 and IGR~J17091--3624 as compared with H1743--322, we
conclude that the mHz QPOS in H1743--322 most probably represent a
different phenomenon than the ``heartbeat'' QPOs and the other
highly-structured low-frequency variability seen in GRS~1915+105 and
IGR~J17091--3624.

%
%

QPOs with intensity-independent frequencies in the mHz range have been
found in at least four NS systems
\citep{Revnivtsev01,Strohmayer11}. The occurrence of these QPOs
depends on source state, but they are thought to be the signature of
marginally stable burning of helium on the NS surface \citep{Heger07}.
A similar QPO, but with an intensity-dependent frequency was also
found in the 11 Hz X-ray pulsar IGR J17480--2446 in Terzan 5
\citep{Linares11}.
The fact that (i) the occurrence of these NS oscillations
are intimately related with thermonuclear X-ray bursts, (ii) are
thought to come from the NS surface and (iii) their spectrum
is generally soft \citep{Revnivtsev01,Altamirano08c}, indicates that
they are most probably a different phenomena from what we detect in
H1743--322.

%
%

\citet{Wijnands04} reported a modulation at $\sim$1~Hz in the light
curve of the accreting millisecond X-ray pulsar (AMXP)
SAX~J1808.4--3658. A similar type of QPO \citep{Patruno10b} was found
in the AMXP NGC~6440 X-2 \citep{Altamirano09}. These QPOs have been
seen at low luminosities (less than a few $10^{36}$ erg s$^{-1}$), at
frequencies between 0.8 and 1.6 Hz and with large amplitudes (up to
100\% fractional rms  \citep{Patruno09d}).
The high amplitude of theses oscillations, the fact that they have
only been seen in AMXPs, and that their occurrence is most probably
related to the onset of the propeller
regime \citep{Patruno09d}
suggests that these $\sim$1~Hz QPOs are not related to those we see in
H1743--322.

%
%

A so called ``1~Hz QPO'' has been reported for (four) dipping NS
systems \citep{Jonker99, Jonker00, Homan03a,Bhattacharyya06c}.  These
QPOs appear to be different from the ``zoo'' of correlated
low-frequency features seen in the power spectra of NS systems.
The fractional rms amplitude of these QPOs is approximately constant
and energy independent during the persistent emission, dips and
thermonuclear X-ray bursts. Although the QPO frequency has been seen
to vary between 0.6 and 2.4 Hz in two of the sources \citep{Jonker00,
  Homan99}, the ``1~Hz QPO'' name stands for the fact that its
frequency can be rather constant for long periods of time.
It has been suggested that these QPOs are related only to high
inclination sources from which we might be observing modulation
effects of the accretion stream material falling onto the disk, or
some kind of modulation produced at the disk edge \citep[e.g.,][and
  references within]{Jonker00,Smale01,Vanderklis06}.
The fact that H1743--322 is thought to be a high inclination source
\citep{Homan05b}, that the fractional rms amplitude of the mHz QPOs we
find does not vary strongly with energy and that its frequency is
stable, indicate that they might be related to the process that
produces the so called ``1~Hz QPO'' in dipping NS systems. If true,
this would be the first BH system showing such QPOs, raising the
question of why the frequency of the QPO is between one and two orders
of magnitude lower in H1743--322 than in the NS.
One possibility is that the frequency range in which this QPO occurs
scales with mass or that it depends on the orbital period of the system,
as H1743--322 is thought to have an orbital period longer than 10 hr
\citep{Jonker10} while the NS systems have orbital periods shorter
than 6 hr \citep[e.g.,][]{Jonker99, Jonker00, Homan03a}.
It is worth noting that the high inclination dipping BH 4U~1630--47
shows QPOs with frequency as low as $\sim$0.1~Hz, however, these QPOs
are due to semi-regular short ($\sim$5 sec) dips \citep{Dieters00}.
Following \citet{Kuulkers98} we produced colors using 1 sec light
curves in different bands of the 2010 and 2011 observation of
H1743--322. We do not observe any hardening (or any other variation)
of the spectra as a function of QPO phase, implying that the 11 mHz
QPOs are most probably not regular dips as seen in 4U~1630--47.

%
%

H1743--322 is well known for its radio jets \citep[e.g.,][]{Steiner11,
Miller-Jones12}. \citet{Markoff05} have suggested that the hard
state emission could be due to synchrotron self-Compton emission from
the base of the jet, and \citet{Russell10} have recently suggested that the
jet mechanism might have dominated the X-ray variability in a portion of
the hard state of the 2000 outburst of the BHC XTE~J1550--556.
The 11 mHz QPOs we find in H1743--322 appear to be different from most
types of variability seen in other BH and NS systems, and so we
speculate whether these oscillations could be related to the radio
jets observed in H1743--322.
Unfortunately no radio measurements have been reported as yet for the
2010 and 2011 outbursts of H1743--322 \citep[and the radio flux is
  known to change between outbursts,][]{Miller-Jones12} and to our
knowledge no model yet predicts that the jet could sometimes modulate
the X-ray flux at a characteristic frequency.
Clearly more theoretical work is needed in this direction.
If related to the jets, it remains unexplained why these low-frequency
QPOs have not yet been identified in other BHs with known radio
emission. Clearly, a systematic search for this type of low-frequency
QPO in the RXTE BH archive is needed to test this speculation.



\begin{thebibliography}{48}
\expandafter\ifx\csname natexlab\endcsname\relax\def\natexlab#1{#1}\fi
\expandafter\ifx\csname url\endcsname\relax
  \def\url#1{{\tt #1}}\fi
\expandafter\ifx\csname urlprefix\endcsname\relax\def\urlprefix{URL }\fi

\bibitem[{{Altamirano} et~al.(2008{\natexlab{a}}){Altamirano}, {van der Klis},
  {M{\'e}ndez} et~al.}]{Altamirano08}
{Altamirano} D., {van der Klis} M., et~al., 2008{\natexlab{a}}, \apj, 685, 436

\bibitem[{{Altamirano} et~al.(2008{\natexlab{b}}){Altamirano}, {van der Klis},
  {Wijnands}, \& {Cumming}}]{Altamirano08c}
{Altamirano} D., {van der Klis} M., et~al., 2008{\natexlab{b}}, \apjl, 673, L35

\bibitem[{{Altamirano} et~al.(2009){Altamirano}, {Strohmayer}, {Heinke}
  et~al.}]{Altamirano09}
{Altamirano} D., {Strohmayer} T.E., et~al., 2009, The Astronomer's Telegram,
  2182, 1

\bibitem[{{Altamirano} et~al.(2011){Altamirano}, {Belloni}, {Linares}
  et~al.}]{Altamirano11d}
{Altamirano} D., {Belloni} T., et~al., Dec. 2011, \apjl, 742, L17

\bibitem[{{Barthelmy} et~al.(2005){Barthelmy}, {Barbier}, {Cummings}
  et~al.}]{Barthelmy05}
{Barthelmy} S.D., {Barbier} L.M., et~al., 2005, Space Science Reviews, 120, 143

\bibitem[{{Belloni} et~al.(2000){Belloni}, {Klein-Wolt}, {M{\'e}ndez}, {van der
  Klis}, \& {van Paradijs}}]{Belloni00}
{Belloni} T., {Klein-Wolt} M., et~al., 2000, \aap, 355, 271

\bibitem[{{Belloni} et~al.(2002){Belloni}, {Psaltis}, \& {van der
  Klis}}]{Belloni02}
{Belloni} T., {Psaltis} D., {van der Klis} M., 2002, \apj, 572, 392

\bibitem[{{Bhattacharyya} et~al.(2006){Bhattacharyya}, {Strohmayer},
  {Markwardt}, \& {Swank}}]{Bhattacharyya06c}
{Bhattacharyya} S., {Strohmayer} T.E., et~al., Mar. 2006, \apjl, 639, L31

\bibitem[{{Casella} et~al.(2004){Casella}, {Belloni}, {Homan}, \&
  {Stella}}]{Casella04}
{Casella} P., {Belloni} T., et~al., 2004, \aap, 426, 587

\bibitem[{{Casella} et~al.(2005){Casella}, {Belloni}, \& {Stella}}]{Casella05}
{Casella} P., {Belloni} T., {Stella} L., 2005, \apj, 629, 403

\bibitem[{{Dieters} et~al.(2000){Dieters}, {Belloni}, {Kuulkers}
  et~al.}]{Dieters00}
{Dieters} S.W., {Belloni} T., et~al., Jul. 2000, \apj, 538, 307

\bibitem[{{Done} et~al.(2004){Done}, {Wardzi{\'n}ski}, \&
  {Gierli{\'n}ski}}]{Done04}
{Done} C., {Wardzi{\'n}ski} G., {Gierli{\'n}ski} M., Apr. 2004, \mnras, 349,
  393

\bibitem[{{Heger} et~al.(2007){Heger}, {Cumming}, \& {Woosley}}]{Heger07}
{Heger} A., {Cumming} A., {Woosley} S.E., 2007, \apj, 665, 1311

\bibitem[{{Homan} et~al.(1999){Homan}, {Jonker}, {Wijnands}, {van der Klis}, \&
  {van Paradijs}}]{Homan99}
{Homan} J., {Jonker} P.G., et~al., 1999, \apjl, 516, L91

\bibitem[{{Homan} et~al.(2003){Homan}, {Wijnands}, \& {van den
  Berg}}]{Homan03a}
{Homan} J., {Wijnands} R., {van den Berg} M., 2003, \aap, 412, 799

\bibitem[{{Homan} et~al.(2005){Homan}, {Miller}, {Wijnands} et~al.}]{Homan05b}
{Homan} J., {Miller} J.M., et~al., 2005, \apj, 623, 383

\bibitem[{{Jahoda} et~al.(2006){Jahoda}, {Markwardt}, {Radeva}
  et~al.}]{Jahoda06}
{Jahoda} K., {Markwardt} C.B., et~al., 2006, \apjs, 163, 401

\bibitem[{{Jonker} et~al.(1999){Jonker}, {van der Klis}, \&
  {Wijnands}}]{Jonker99}
{Jonker} P.G., {van der Klis} M., {Wijnands} R., 1999, \apjl, 511, L41

\bibitem[{{Jonker} et~al.(2000){Jonker}, {M{\' e}ndez}, \& {van der
  Klis}}]{Jonker00}
{Jonker} P.G., {M{\' e}ndez} M., {van der Klis} M., 2000, \apjl, 540, L29

\bibitem[{{Jonker} et~al.(2010){Jonker}, {Miller-Jones}, {Homan}
  et~al.}]{Jonker10}
{Jonker} P.G., {Miller-Jones} J., et~al., Jan. 2010, \mnras, 401, 1255

\bibitem[{{Kaluzienski} \& {Holt}(1977)}]{Kaluzienski77}
{Kaluzienski} L.J., {Holt} S.S., Aug. 1977, \iaucirc, 3099, 3

\bibitem[{{Klein-Wolt}(2004)}]{Kleinwolt04}
{Klein-Wolt} M., 2004, PhD.Thesis

\bibitem[{{Kuulkers} et~al.(1998){Kuulkers}, {Wijnands}, {Belloni}
  et~al.}]{Kuulkers98}
{Kuulkers} E., {Wijnands} R., et~al., 1998, \apj, 494, 753

\bibitem[{{Linares} et~al.(2011){Linares}, {Chakrabarty}, \& {van der
  Klis}}]{Linares11}
{Linares} M., {Chakrabarty} D., {van der Klis} M., Feb. 2011, ArXiv e-prints

\bibitem[{{Lomb}(1976)}]{Lomb76}
{Lomb} N.R., 1976, \apss, 39, 447

\bibitem[{{Markoff} et~al.(2005){Markoff}, {Nowak}, \& {Wilms}}]{Markoff05}
{Markoff} S., {Nowak} M.A., {Wilms} J., Dec. 2005, \apj, 635, 1203

\bibitem[{{McClintock} et~al.(2009){McClintock}, {Remillard}, {Rupen}
  et~al.}]{McClintock09}
{McClintock} J.E., {Remillard} R.A., et~al., Jun. 2009, \apj, 698, 1398

\bibitem[{{M{\'e}ndez} et~al.(1999){M{\'e}ndez}, {van der Klis}, {Ford},
  {Wijnands}, \& {van Paradijs}}]{Mendez99a}
{M{\'e}ndez} M., {van der Klis} M., et~al., 1999, \apjl, 511, L49

\bibitem[{{Miller-Jones} et~al.(2012){Miller-Jones}, {Sivakoff}, {Altamirano}
  et~al.}]{Miller-Jones12}
{Miller-Jones} J.C.A., {Sivakoff} G.R., et~al., Jan. 2012, ArXiv e-prints

\bibitem[{{Miyamoto} et~al.(1993){Miyamoto}, {Iga}, {Kitamoto}, \&
  {Kamado}}]{Miyamoto93}
{Miyamoto} S., {Iga} S., et~al., 1993, \apjl, 403, L39

\bibitem[{{Motta} et~al.(2010){Motta}, {Mu{\~n}oz-Darias}, \&
  {Belloni}}]{Motta10}
{Motta} S., {Mu{\~n}oz-Darias} T., {Belloni} T., Nov. 2010, \mnras, 408, 1796

\bibitem[{{Neilsen} et~al.(2011){Neilsen}, {Remillard}, \& {Lee}}]{Neilsen11}
{Neilsen} J., {Remillard} R.A., {Lee} J.C., Aug. 2011, \apj, 737, 69

\bibitem[{{Patruno} et~al.(2009){Patruno}, {Watts}, {Klein Wolt}, {Wijnands},
  \& {van der Klis}}]{Patruno09d}
{Patruno} A., {Watts} A., et~al., 2009, \apj, 707, 1296

\bibitem[{{Patruno} et~al.(2010){Patruno}, {Yang}, {Altamirano}
  et~al.}]{Patruno10b}
{Patruno} A., {Yang} Y., et~al., Jun. 2010, The Astronomer's Telegram, 2672, 1

\bibitem[{{Press et al.}(1992)}]{Numerical-Recipes}
{Press et al.}, 1992, Numerical Recipes: The Art of Scientific Computing,
  Cambridge University Press, Cambridge (UK) and New York, 2nd edn.

\bibitem[{{Revnivtsev} et~al.(2001){Revnivtsev}, {Churazov}, {Gilfanov}, \&
  {Sunyaev}}]{Revnivtsev01}
{Revnivtsev} M., {Churazov} E., et~al., 2001, \aap, 372, 138

\bibitem[{{Revnivtsev} et~al.(2003){Revnivtsev}, {Chernyakova}, {Capitanio}
  et~al.}]{Revnivtsev03b}
{Revnivtsev} M., {Chernyakova} M., et~al., Mar. 2003, The Astronomer's
  Telegram, 132, 1

\bibitem[{{Russell} et~al.(2010){Russell}, {Maitra}, {Dunn}, \&
  {Markoff}}]{Russell10}
{Russell} D.M., {Maitra} D., et~al., Jul. 2010, \mnras, 405, 1759

\bibitem[{{Scargle}(1982)}]{Scargle82}
{Scargle} J.D., 1982, \apj, 263, 835

\bibitem[{{Smale} et~al.(2001){Smale}, {Church}, \&
  {Ba{\l}uci{\'n}ska-Church}}]{Smale01}
{Smale} A.P., {Church} M.J., {Ba{\l}uci{\'n}ska-Church} M., 2001, \apj, 550,
  962

\bibitem[{{Steiner} et~al.(2011){Steiner}, {McClintock}, \& {Reid}}]{Steiner11}
{Steiner} J.F., {McClintock} J.E., {Reid} M.J., Nov. 2011, ArXiv e-prints

\bibitem[{{Strohmayer}(2011)}]{Strohmayer11a}
{Strohmayer} T.E., Apr. 2011, The Astronomer's Telegram, 3277, 1

\bibitem[{{Strohmayer} \& {Smith}(2011)}]{Strohmayer11}
{Strohmayer} T.E., {Smith} E.A., Apr. 2011, The Astronomer's Telegram, 3258, 1

\bibitem[{{van der Klis}(1994)}]{Vanderklis94}
{van der Klis} M., Jun. 1994, \apjs, 92, 511

\bibitem[{{van der Klis}(1995)}]{Vanderklis95b}
{van der Klis} M., 1995, Proceedings of the NATO Advanced Study Institute on
  the Lives of the Neutron Stars, held in Kemer, Turkey, August 19-September
  12, 1993. Editor(s), M. A. Alpar, U. Kiziloglu, J. van Paradijs; Publisher,
  Kluwer Academic, Dordrecht, The Netherlands, Boston, Massachusetts, 301

\bibitem[{{van der Klis}(2006)}]{Vanderklis06}
{van der Klis} M., 2006, in Compact Stellar X-Ray Sources, ed. W. H. G. Lewin
  \& M. van der Klis (Cambridge: Cambridge Univ. Press)

\bibitem[{{Wijnands}(2004)}]{Wijnands04}
{Wijnands} R., Jun. 2004, Nuclear Physics B Proceedings Supplements, 132, 496

\bibitem[{{Wijnands} \& {van der Klis}(1999)}]{Wijnands99a}
{Wijnands} R., {van der Klis} M., 1999, \apj, 514, 939

\end{thebibliography}
\end{document}